\documentclass[preprint,prd,showpacs,floatfix]{revtex4}

\usepackage{amsmath}
\usepackage{amsfonts}
\usepackage{amssymb}
\usepackage{bbm}
\usepackage{latexsym}
\usepackage[dvips]{graphicx}
\usepackage{graphics}
\usepackage{color}

\begin{document}

\title{Toward reconstruction of dynamics of the Universe from distant supernovae type Ia } 
\author{Marek Szyd{\l}owski}
\email{uoszydlo@cyf-kr.edu.pl}
\affiliation{Astronomical Observatory, Jagiellonian University, Krak\'ow, Poland}
\author{Wojciech Czaja}
\email{czaja@oa.uj.edu.pl}
\affiliation{Astronomical Observatory, Jagiellonian University, Krak\'ow, Poland}
\date{\today}

\begin{abstract}
We demonstrate a model-independent method of estimating qualitative dynamics of an accelerating universe
from observations of distant type Ia supernovae. Our method is based on the luminosity-distance 
function, optimized to fit observed distances of supernovae, and the Hamiltonian representation of dynamics
for the quintessential universe with a general form of equation of state $p=w(a(z))\rho$. 
Because of the Hamiltonian structure of FRW dynamics with the equation of state $p = w(a(z)) \rho$, the 
dynamics is uniquelly determined by the potential function $V(a)$ of the system. The effectiveness
of this method in discrimination of model parameters of Cardassian evolution scenario is also given.
Our main result is the following, restricting to the flat model with the current value of $\Omega_{m,0}=0.3$,
the constraints at $2\sigma$ confidence level to the presence of $\rho^{n}$ modification of the FRW models 
are $-0.50 \lesssim n \lesssim 0.36$.
\end{abstract}


\maketitle

\section{Introduction}

Recent observations of type Ia supernovae \cite{riess98,riess02,perlmutter99} supported by WMAP measurements
of anisotropy of the angular temperature fluctuations \cite{benoit03} indicate that our Universe is 
spatially flat and accelerating. On the other hand, the power spectrum of galaxy clustering \cite{parcival02}
indicates that about $30\%$ of critical density of the universe should be in the form of non-relativistic
matter (cold dark matter and baryons). The remaining, almost two thirds of the critical energy, may be
in the form of a component having negative pressure (dark energy). Although the nature of dark energy
is unknown, the positive cosmological constant term seems to be a serious candidate for the 
description of dark energy. In this case the cosmological constant $\Lambda$ and energy density 
$\varepsilon_{\Lambda}=\Lambda/(8\pi G) $ remain constant with time and the corresponding mass density
$\rho_{\Lambda} \equiv \varepsilon_{\Lambda}/c^{2} = 6.44 \times 10^{-30}(\Omega_{\Lambda}/0.7)(h/0.7) g$ 
$cm^{-3}$, where $h$ is the Hubble constant $H_{0}$ expressed in units of $100$ ${\rm km}$ 
${\rm s^{-1} Mpc^{-1}}$ and $\Omega_{\Lambda} = 0.7 \pm 0.1$. Although the cold dark matter (CDM) model 
with the cosmological constant and dust provides an excellent explanation of the SNIa data, 
the present value of $\Lambda$ is $\sim 10^{123}$ times smaller than value predicted by the particle physics 
model. Many alternative condidates for dark energy have been advanced and some of them are in good 
agreement with the current observational constraints 
\cite{ratra98,peebles03,caldwell98,kamenshchik01,armendariz00,bucher99,parker99}. 
Moreover, it is a natural suggestion that $\Lambda$-term has a dynamical nature like in the inflationary 
scenario. Therefore, it is reasonable to consider the next simplest form of dark energy alternative to the 
cosmological constant $(w=-1)$ for which the equation of state depends upon time in such a way that 
$p = w(a(z))\rho$, where $w \equiv p/ \rho$ is a coefficient of the equation of state parametrized by the 
scale factor or redshift. It has been demonstrated \cite{szydlowski03,szydlowski03a} that dynamics of such 
a system can be represented by one-dimensional Hamiltonian flow
\begin{equation}
\mathcal{H} = \frac{\dot{a}^{2}}{2} + V(a),
\label{eq:1}
\end{equation}
where the overdot means differentiation with respect to the cosmological time $t$ and $V(a)$ is a potential
function of the scale factor $a$ given by
\begin{equation}
V(a) = - \frac{\rho_{\rm eff}(a)}{6}a^{2},
\label{eq:2}
\end{equation}
where $\rho_{\rm eff}$ is the effective energy density which satisfies the conservation condition
\begin{equation*}
\dot{\rho}_{\rm eff} =~-3~\frac{\dot{a}}{a}~(\rho_{\rm eff}+p_{\rm eff}).
\end{equation*}
For example, for the ${\rm \Lambda CDM}$ model we have 
\begin{gather}
\rho_{\rm eff} = \Lambda + \rho_{m,0}a^{-3} \nonumber \\
p_{\rm eff} = - \Lambda + 0.
\label{eq:3}
\end{gather}
Of course the trajectories of the system lie on the zero energy surface $\mathcal{H} \equiv 0$.

Hamiltonian (\ref{eq:1}) can be rewritten in the following form convenient for our current 
reconstruction of the equation of state from the potential function $V$, namelly
\begin{equation}
\mathcal{H}(p,x) = \frac{p_{x}^{2}}{2} + V(x),
\label{eq:4}
\end{equation}
where $p_{x} = \partial L/\partial \dot{x}$, here overdot means differentiation with respect to some new 
reparametrized time $t \rightarrow \tau \colon d \tau = | H_{0} |dt$, $x \equiv a/a_{0} = 1+z$.

For example, for mixture of non-interacting fluids potential $V(x)$ takes the form
\begin{equation}
V(x) = - \frac{1}{2} \sum_{i} \Omega_{i,0} x^{2-3(w_{i} + 1)},
\label{eq:5}
\end{equation}
where $p_{i} = w_{i} \rho_{i}$ for $i$-th fluid and $\forall i$ $w_{i} = {\rm const.}$ (similar to the 
quiessence model of dark energy).

Due to the Hamiltonian structure of Friedmann-Robertson-Walker dynamics, with the general form of the 
equation of state $p = w(a(z)) \rho$, the dynamics is uniquely determined by the potential function $V(a)$ 
(or $V(x)$) of the system. Only for simplicity of presentation we assume that the universe is spatially 
flat (in the opposite case trajectories of the system should be considered on the energy level 
$\mathcal{H} = \frac{1}{2} \Omega_{k,0}$).

Let us note that from the potential function we can obtain the equation of state coefficient 
$w = p/ \rho$,
\begin{equation}
w = - \frac{1}{3} \bigg(1 + \frac{d(\ln{V})}{d(\ln{a})}\bigg).
\label{eq:6}
\end{equation}
The term $d(\ln{V})/d(\ln{a})$ has a simple interpretation as an elasticity coefficient of the potential 
function with respect to the scale factor. 

Thus from the potential function $V$ both $\rho(a)$ and $p$ can be unambiguously calculated
\begin{gather}
\rho(a) = - 6\frac{V(a)}{a^{2}},\nonumber \\
p = 2\frac{V(a)}{a^{2}} \bigg(1+\frac{d(\ln{V})}{d(\ln{a})}\bigg).
\label{eq:7}
\end{gather}

\section{The expansion scenario from supernovae distances}

As it is well known in a flat FRW cosmology the luminosity distance $D_{L}$ and the coordinate distance $r$ 
to an object at redshift $z$ are simply related as 
\begin{equation}
a_{0}r = a_{0}\int_{t}^{t_{0}}\frac{dt'}{a(t')} = \frac{D_{L}(z)}{1+z},
\label{eq:8}
\end{equation}
($c = 8 \pi G = 1$ here and elsewhere).
From equation (\ref{eq:8}) the Hubble parameter is given by
\begin{equation}
H(z) = \dot{[\ln{a}]} = \bigg[\frac{d}{dz}\bigg(\frac{D_{L}(z)}{1+z}\bigg)\bigg].
\label{eq:9}
\end{equation}

It is crucial that formula (\ref{eq:9}) is purely kinematic and depends neither upon a microscopic 
model of matter, including the $\Lambda$-term, nor on a dynamical theory of gravity. Due to existance of 
such a relation it would be possible to calculate the potential function which is:
\begin{equation}
V(a) = - \frac{1}{2}H^{2}a^{2} = 
- \frac{\big[\frac{d}{dz}\big(\frac{D_{L}(z)}{1+z}\big)\big]^{-2}}{2(1+z)^{2}}.
\label{eq:10}
\end{equation}

This in turn allows us to reconstruct the potential $V(z)$ from SNIa data. Let us note that $V(z)$ depends 
on the first derivative with respect to $z$ whereas $w(z)$ is associated with the second derivative. 

Let us also note that from of the potential function for a one-dimensional particle-universe moving in the 
configurational $a$ (or $x$)-space can be reconstructed from recent measurements of angular size of high-z 
compact radio sources compiled by Gurvits {\it et al.} \cite{gurvits99}. The corresponding formula is
\begin{equation}
V(a)=-\frac{\big[\frac{d}{dz}\big(D_{A}(z)(1+z)\big)\big]^{-2}}{2(1+z)^{2}},
\label{eq:10a}
\end{equation}
where the luminosity distance $D_{L}$ and the angular distance $D_{A}$ are related by the simple formula
\begin{equation}
D_{L}(z)=(1+z)^{2}D_{A}(z).
\label{eq:10b}
\end{equation}

Since the potential function is related to the luminosity function by relation (\ref{eq:10}) one can
determine both the class of trajectories in the phase plane $(a,\dot{a})$ and the Hamiltonian form as well 
as reconstruct the quintessence parameter $w(z)$ provided that the luminosity function $D_{L}(z)$ is known 
from observations.

Now we can reconstruct the form of the potential function (\ref{eq:10}) using a natural Ansatz introduced 
by Sahni {\it et al.} \cite{sahni00}. In this approach dark energy density which coincides with
$\rho_{\rm eff}$ is given as a truncated Taylor series with respect to $u = (1+z)$
\begin{equation*}
\rho_{\rm DE} = \rho_{\rm eff} = A_{0} + A_{1}u + A_{2}u^{2}.
\end{equation*}
This leads to
\begin{equation}
H(z) = H_{0}\sqrt{\Omega_{m,0}(1+z)^{3}+A_{0}+A_{1}(1+z)+A_{2}(1+z)^{2}}
\label{eq:11}
\end{equation}
and
\begin{equation}
\frac{D_{L}(z)}{(1+z)} = \frac{1}{H_{0}}\int_{0}^{z}\frac{dz}{\sqrt{\Omega_{m,0}(1+z)^{3}+A_{0}+A_{1}(1+z)+A_{2}(1+z)^{2}}}.
\label{eq:12}
\end{equation}

The values of three parameters $A_{0}$,$A_{1}$,$A_{2}$ can be obtained by applying a standard fitting 
procedure to SNIa observational data based on the maximum likelihood method.

The potential function (\ref{eq:10}) written in terms of $(1+z)$ is
\begin{equation}
V(a(z)) = -\frac{\rho_{\rm eff}a^{2}}{6} = -\frac{1}{2}H_{0}^{2}\big[\Omega_{m,0}(1+z)+A_{0}(1+z)^{-2}+A_{1}(1+z)^{-1}+A_{2}\big]
\label{eq:13}
\end{equation}
or in dimensionless form
\begin{equation}
V(x(z)) = -\frac{1}{2}\big[\Omega_{m,0}(1+z)+\Omega_{\Lambda,0}(1+z)^{-2}+\Omega_{t,0}(1+z)^{-1}+\Omega_{k,0}\big],
\label{eq:14}
\end{equation}
where $A_{i}/(3H_{0}^{2}) = \Omega_{i,0}$, $(i = \Lambda,t,k)$.

Our approach to the reconstruction of dynamics of the model is different from the standard approach in 
which $w(z)$ is determined directly from the luminosity distance formula. It should be stressed out that the
latter approach has an inevitable limitation because the luminosity distance dependence on $w(z)$ is 
obtained through a multiple-integral relation that loses detailed information on $w(z)$ \cite{maor01}. 
In our approach the reconstruction is simpler and more information on $w(z)$ survives (only a single 
integral is required). Our approach is also different from the concept of reconstruction of potential of 
scalar fields considered in the context of quintessence \cite{caldwell98}.

The key steps of our method are the following:

\noindent
{\bf 1)}
we reconstruct the potential function $V(a)$ for the Hamiltonian dynamics of the quintessential universe
from the luminosity distances of supernovas type Ia; \\
\noindent
{\bf 2)}
we draw the best fit curves and confidence levels regions obtained from the statistical analysis of SNIa
data; \\
\noindent
{\bf 3)}
we set the theoretically predicted forms of the potential functions on the confidence levels diagram; \\
\noindent
{\bf 4)}
those theoretical potential which escape from the $2\sigma$ confidence level is treated as being unfitted 
to observations; \\
{\bf 5)}
we choose this potential function which lie near the best fit curve. \\

Our reconstruction is an effective statistical technique which can be used to compare a large number 
of theoretical models with observations. Instead of estimating some revelant parameters for each model 
separately, we choose a model-independent fitting function and perform a maximum likelihood parameter 
estimation for it. The obtained confidence levels can be used to discriminate between the considered models. 
In this paper this technique is used to find the fitting function for the luminosity distance. 

The additional argument which is important when considering the potential $V(a)$ is that it allows to find 
some modification in the Friedmann equations along the ``Cardassian expansion scenario'' \cite{freese02}. 
This proposition is very intriguing because of additional terms, which automatically cause the acceleration
of the universe \cite{alcaniz99,avelino03,zhu02,zhu03}. These modifications come from the 
fundamental physics and these terms can be tested using astronomical observation of distant type Ia 
supernovae. For this aim the recent measurements of angular size of high-redshift compact radio sources can 
also be used \cite{gurvits99}.

The important question is the reliable data available. We expect that supernovae data would improve greatly 
over next few years. The ongoing SNAP mission should gives us about 2000 type Ia supernovae cases each year. 
This satellite mission and the next planned ones will increase the accuracy of data compared to data from 
the 90s. In our analysis we use the availale data starting from the three Perlmutter samples 
(Sample A is the complete sample of 60 supernovae, but in the analysis it is also used sample B and C in 
which 4 and 6 outliers were excluded, respectively). The fit for the sample C is more robust and this sample
was accepted as the base of our consideration. For technical details of the metod the reader is referred to
our previous two papers \cite{szydlowski03,szydlowski03a}. 

In Fig.~1 we show the reconstructed potential function obtained using the fitting values of $A_{i}$ as well 
as $\mathcal{M}$. The red line represents the potential function for the best fit values of parameters 
(see TABLES~\ref{resultsa}, \ref{resultsc}~and~\ref{resultspac}). In each case the coloured areas cover the 
confidence levels $68.2\%$ ($1\sigma$) and $95.4\%$ ($2\sigma$) for the potential function.
The different forms of the potential function which are obtained from the theory are presented in
the confidence levels. Here we consider one case, namely the Cardassian model.
In this case the standard FRW equation is modified by the presence of an additional $\rho^{n}$ term,
where $\rho$ is the energy density of matter and radiation. For simplicity we assume that density parameter
for radiation is zero (see TABLE.~\ref{pottab}).
The Cardassian scenario is proposed as an alternative to the cosmological constant in explaining the 
acceleration of the Universe. In this scenario the the Universe automaticaly accelerates without any dark 
energy component. The additional term in the Friedmann equation arises from exotic physics of the
early universe (i.e., in the brane cosmology with Randall-Sundrum version $n=2$).
\begin{table}[!ht]
\caption{The forms of the potential functions in dimensionless form for two cases: ${\rm \Lambda CDM}$
model and Cardassian scenario.
}
\begin{center}
\scriptsize
\begin{tabular}{c|c|c|c|c}
\hline
\hline
 & & &\multicolumn{2}{c}{} \\
\parbox{3cm}{}& \parbox{3.5cm}{ The form of the potential function} & \parbox{3.5cm}{Position of the maximum predicted by theorethical models} &\multicolumn{2}{c}{\parbox{3.5cm}{Position of the maximum from reconstruction}} \\
 & & &\multicolumn{2}{c}{} \\
\cline{4-5}
 & & & sample & value \\
\hline
 & & & A & $0.613$ \\
Perlmutter model & $V(x)=-\frac{1}{2}\big[\Omega_{m,0}x^{-1}+\Omega_{\Lambda,0}x^{2}\big]$ & $x_{0}=\sqrt[3]{\frac{1}{2}\frac{\Omega_{m,0}}{1-\Omega_{m,0}}}$ & C & $0.594$ \\
\hline
 & & & A & $0.598$ \\
Cardassian scenario & $V(x)=-\frac{1}{2}\big[\Omega_{m,0}x^{-1}+\Omega_{C,0}x^{2-3n}\big]$ & $x_{0}=\Big[(2-3n)\Big(\frac{1}{\Omega_{m,0}}-1\Big)\Big]^{\frac{1}{3(n-1)}}$ & C & $0.599$ \\
\hline
\end{tabular}
\end{center}
\label{pottab}
\end{table}

\begin{figure}[!ht]
\begin{center}
$\begin{array}{c@{\hspace{0.2in}}c}
\multicolumn{1}{l}{\mbox{\bf a)}} & 
\multicolumn{1}{l}{\mbox{\bf b)}} \\ [-0.5cm]
\includegraphics[scale=0.3, angle=270]{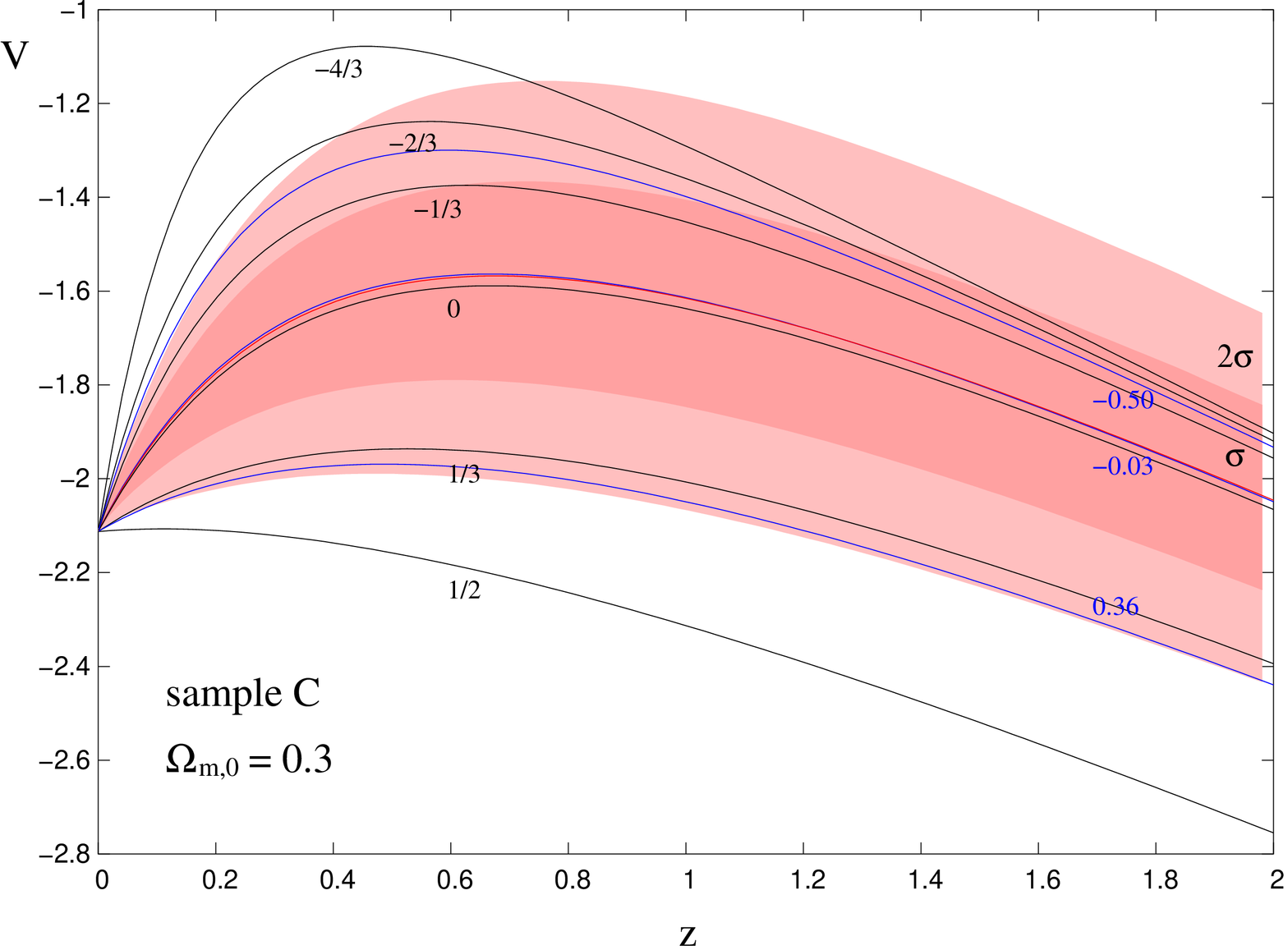} & 
\includegraphics[scale=0.3, angle=270]{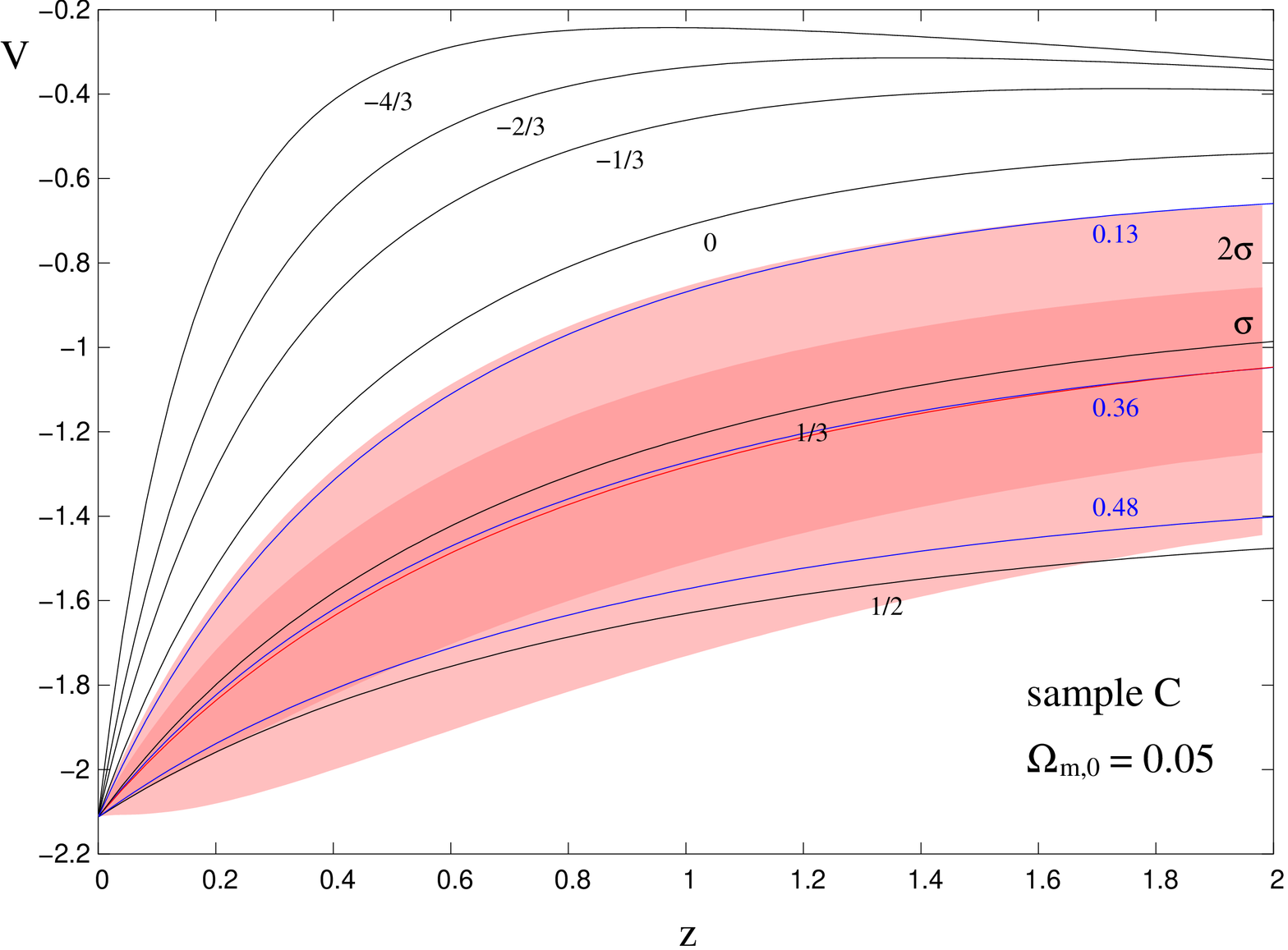} \\ [0.4cm]
\multicolumn{1}{l}{\mbox{\bf c)}} & 
\multicolumn{1}{l}{\mbox{\bf d)}} \\ [-0.5cm]
\includegraphics[scale=0.3, angle=270]{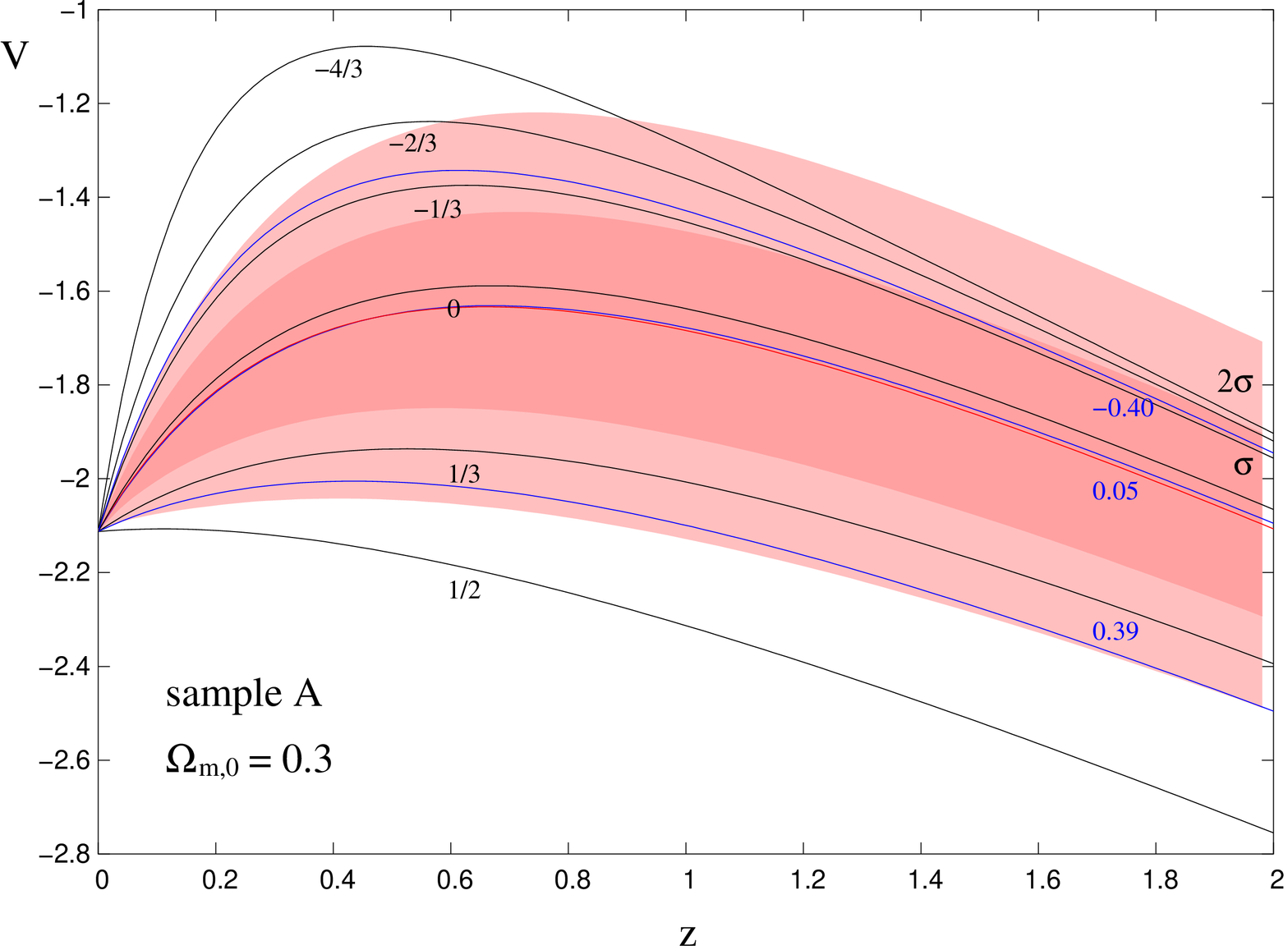} & 
\includegraphics[scale=0.3, angle=270]{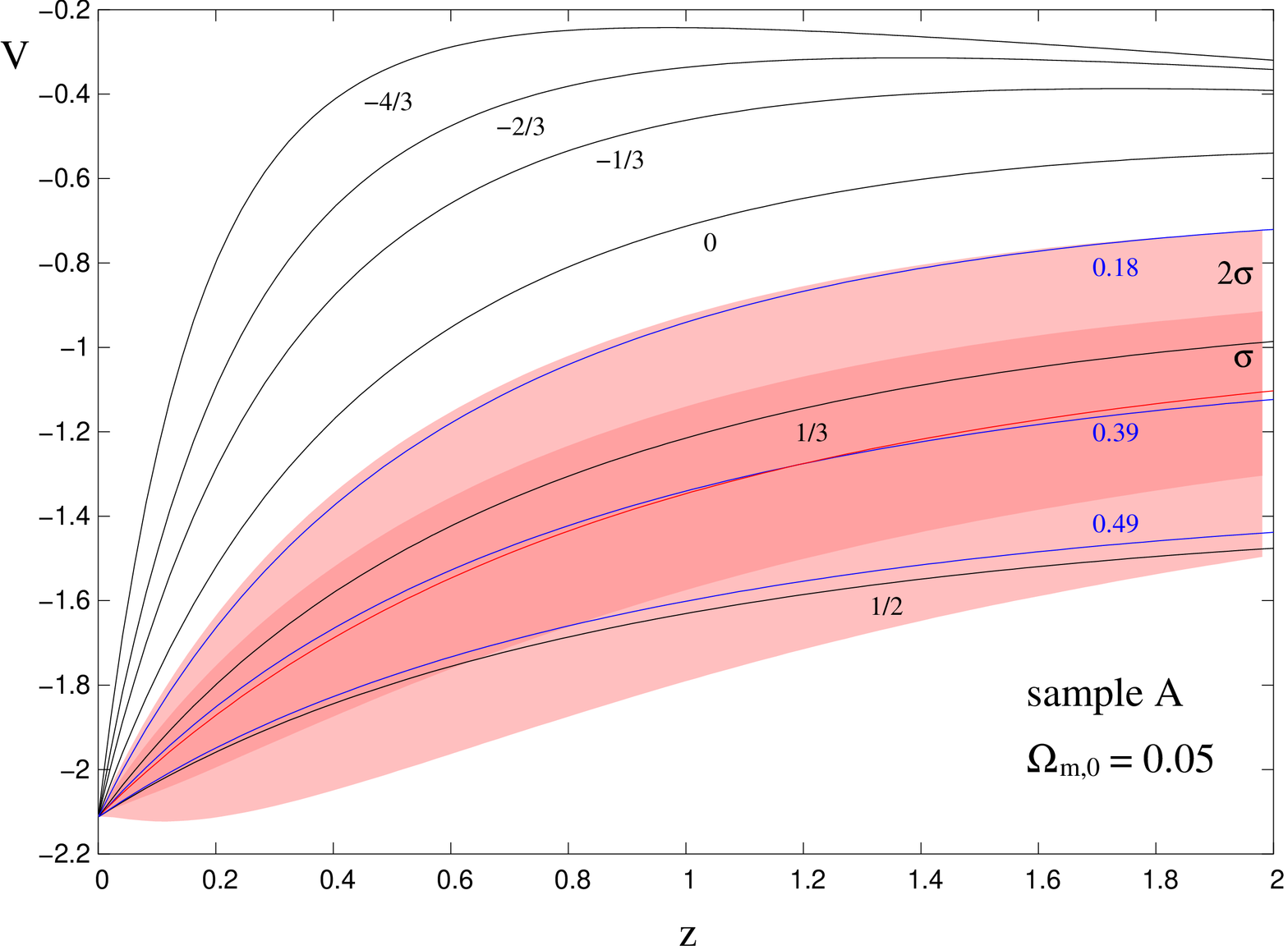} \\ [0.4cm]
\multicolumn{1}{l}{\mbox{\bf e)}} & 
\multicolumn{1}{l}{\mbox{\bf f)}} \\ [-0.5cm]
\includegraphics[scale=0.3, angle=270]{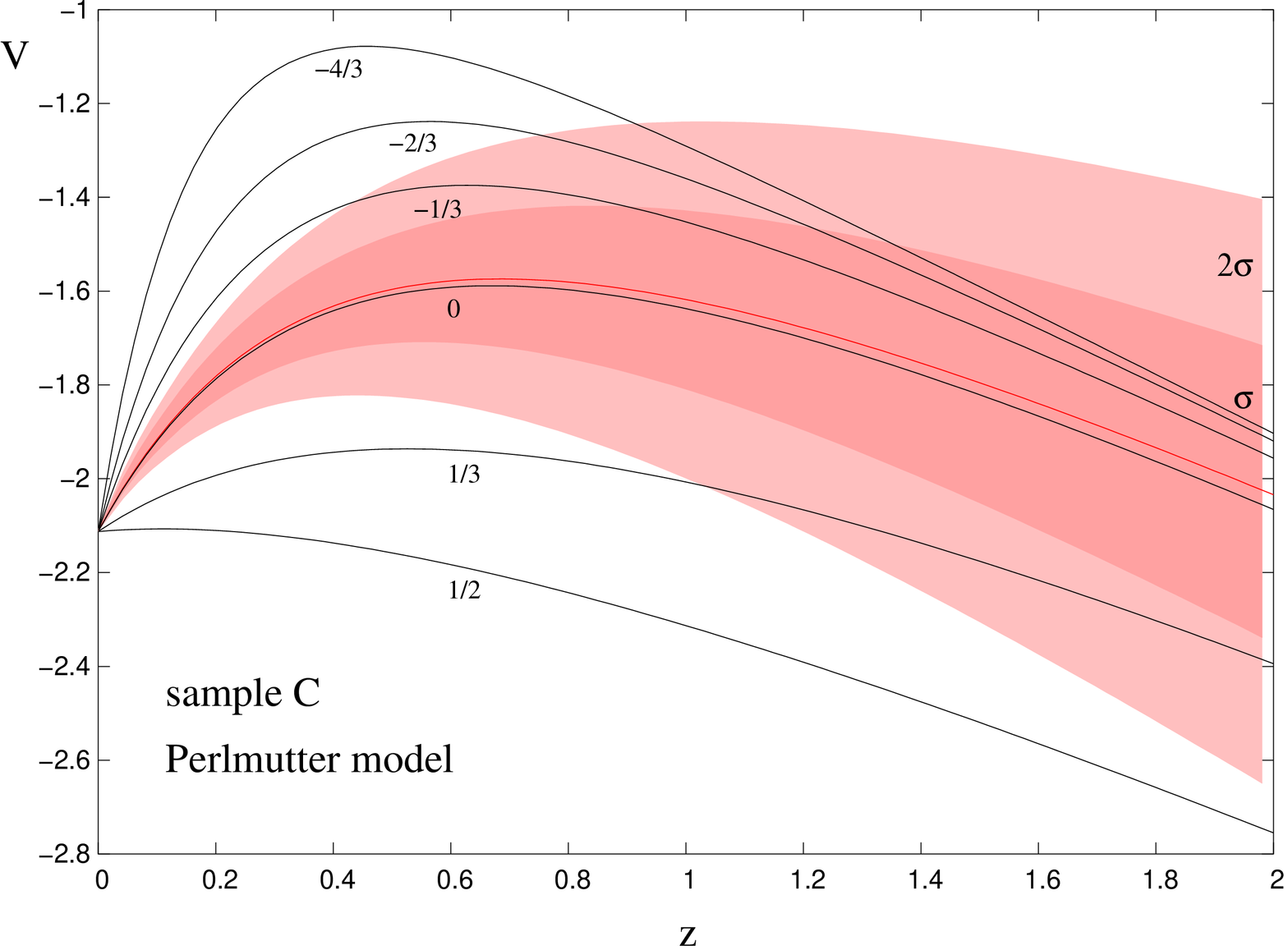} & 
\includegraphics[scale=0.3, angle=270]{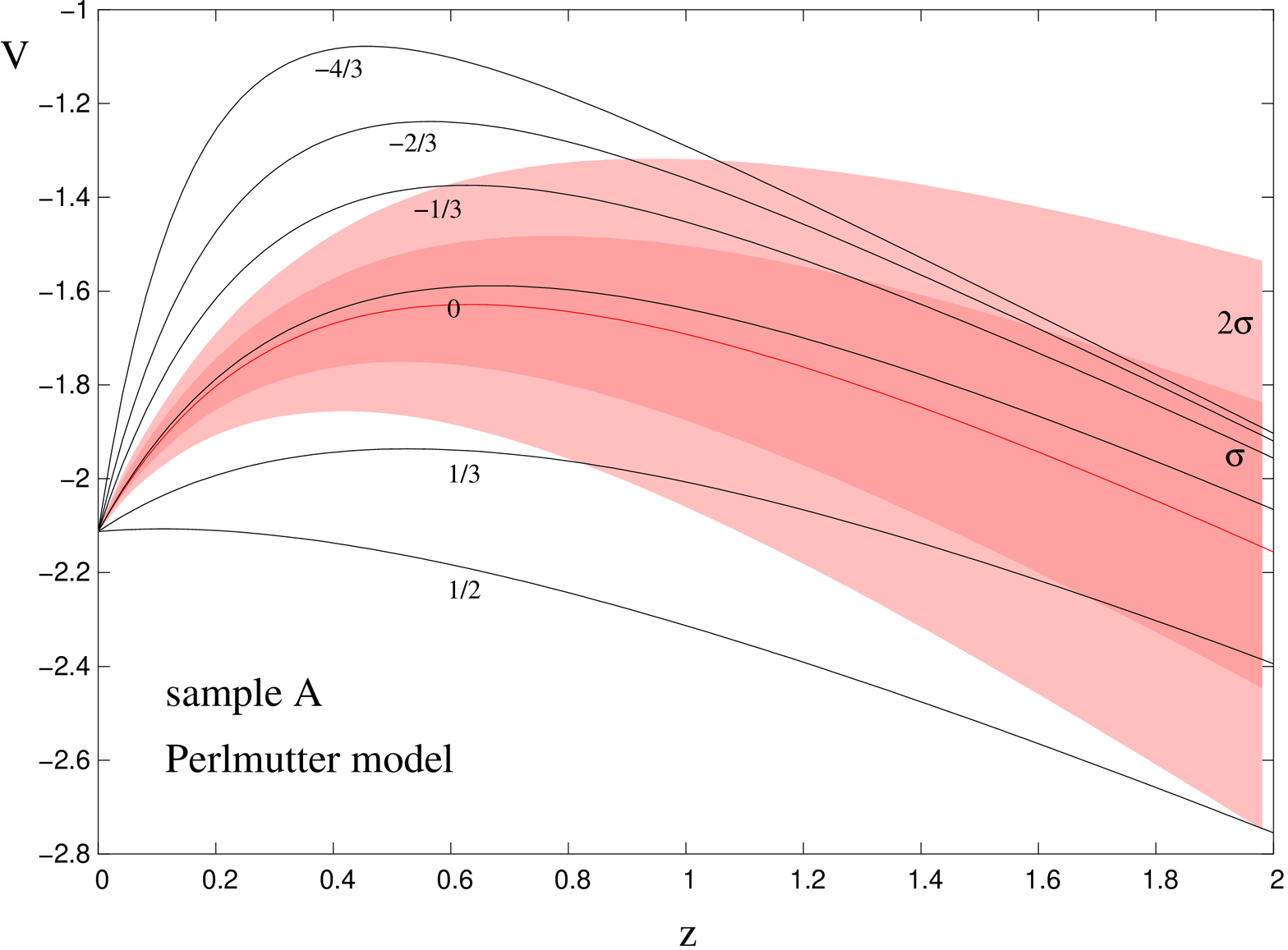} \\ [0.4cm]
\end{array}$
\end{center}
\linespread{0.5}
\caption{
Confidence levels for the potential function $V$ for the model with {\bf a)}  $A_{3}=0.3$, (sample C),
{\bf b)} $A_{3}=0.05$, (sample C), {\bf c)} $A_{3}=0.3$, (sample A), {\bf d)} $A_{3}=0.05$, (sample A);
and for the Perlmutter model {\bf e)} (sample C), {\bf f)} (sample A).
The best fit for each model is represented by the red curve. Black and blue curves represent potential 
functions for theoretical Cardassian models for $\Omega_{m,0}=0.3$ ({\bf a},{\bf b},{\bf e},{\bf f}),
$\Omega_{m,0}=0.05$ ({\bf c},{\bf d}) and $n = 1/2,1/3,0,-1/3,-2/3,-4/3$. 
}
\label{figpotw}
\end{figure}

The solid lines in Fig.~\ref{figpotw} represent the potential functions for the different Cardassian 
scenarios of an accelerated expansion of the universe.

\begin{figure}[!ht]
\begin{center}
$\begin{array}{c@{\hspace{0.2in}}c}
\multicolumn{1}{l}{\mbox{\bf a)}} & 
\multicolumn{1}{l}{\mbox{\bf b)}} \\ [-0.5cm]
\includegraphics[scale=0.3, angle=270]{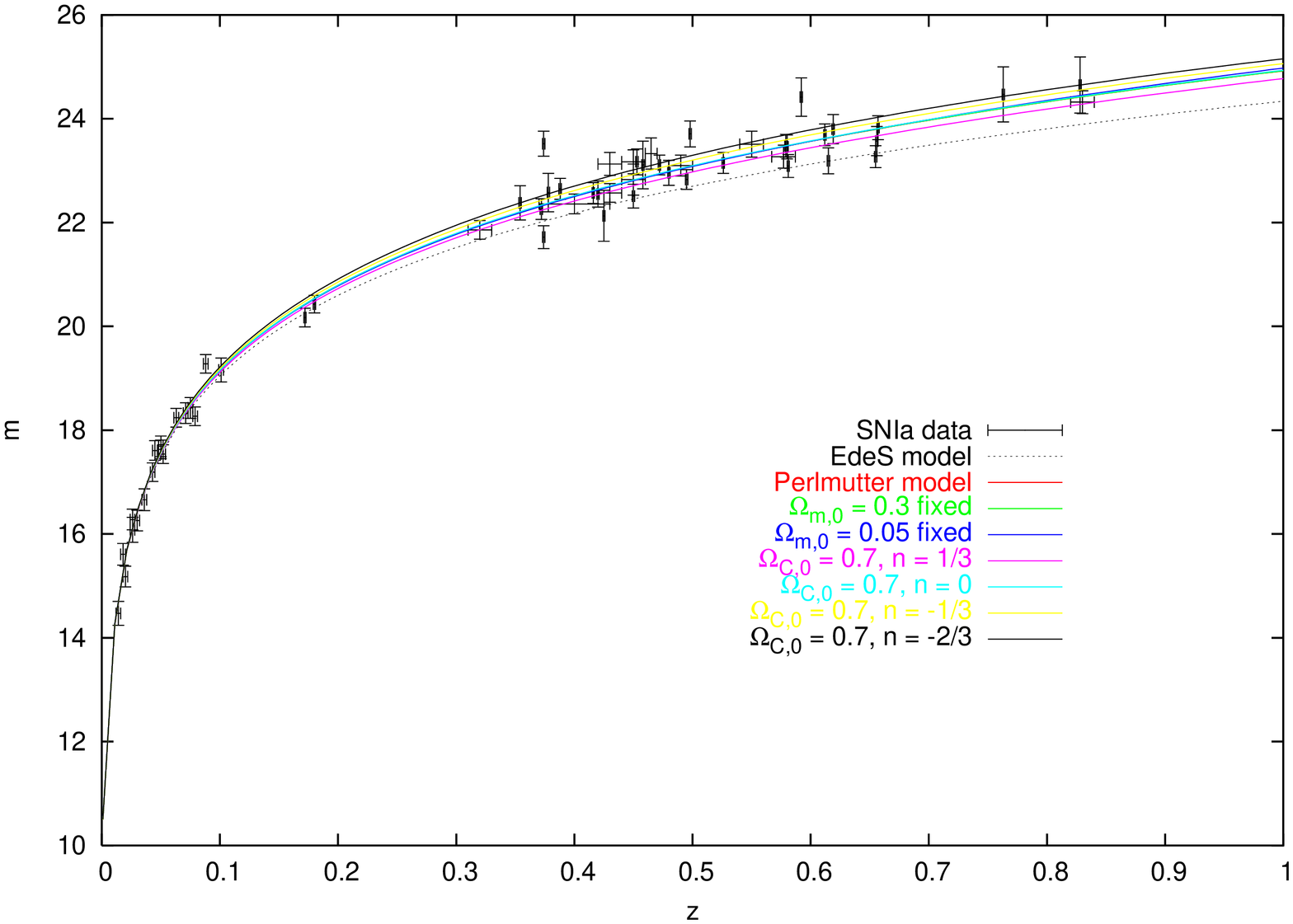} & 
\includegraphics[scale=0.3, angle=270]{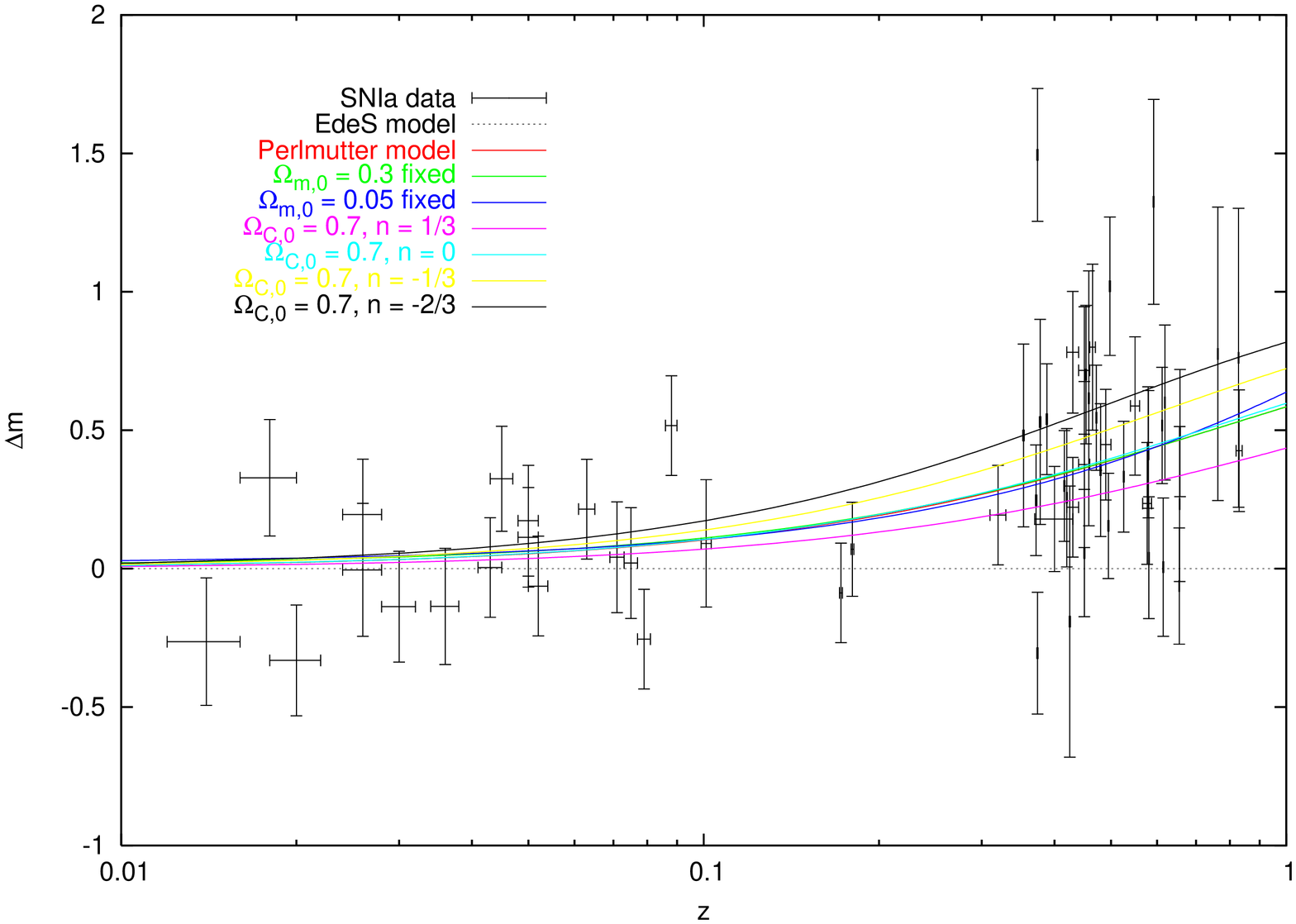} \\ [0.4cm]
\multicolumn{1}{l}{\mbox{\bf c)}} & 
\multicolumn{1}{l}{\mbox{\bf d)}} \\ [-0.5cm]
\includegraphics[scale=0.3, angle=270]{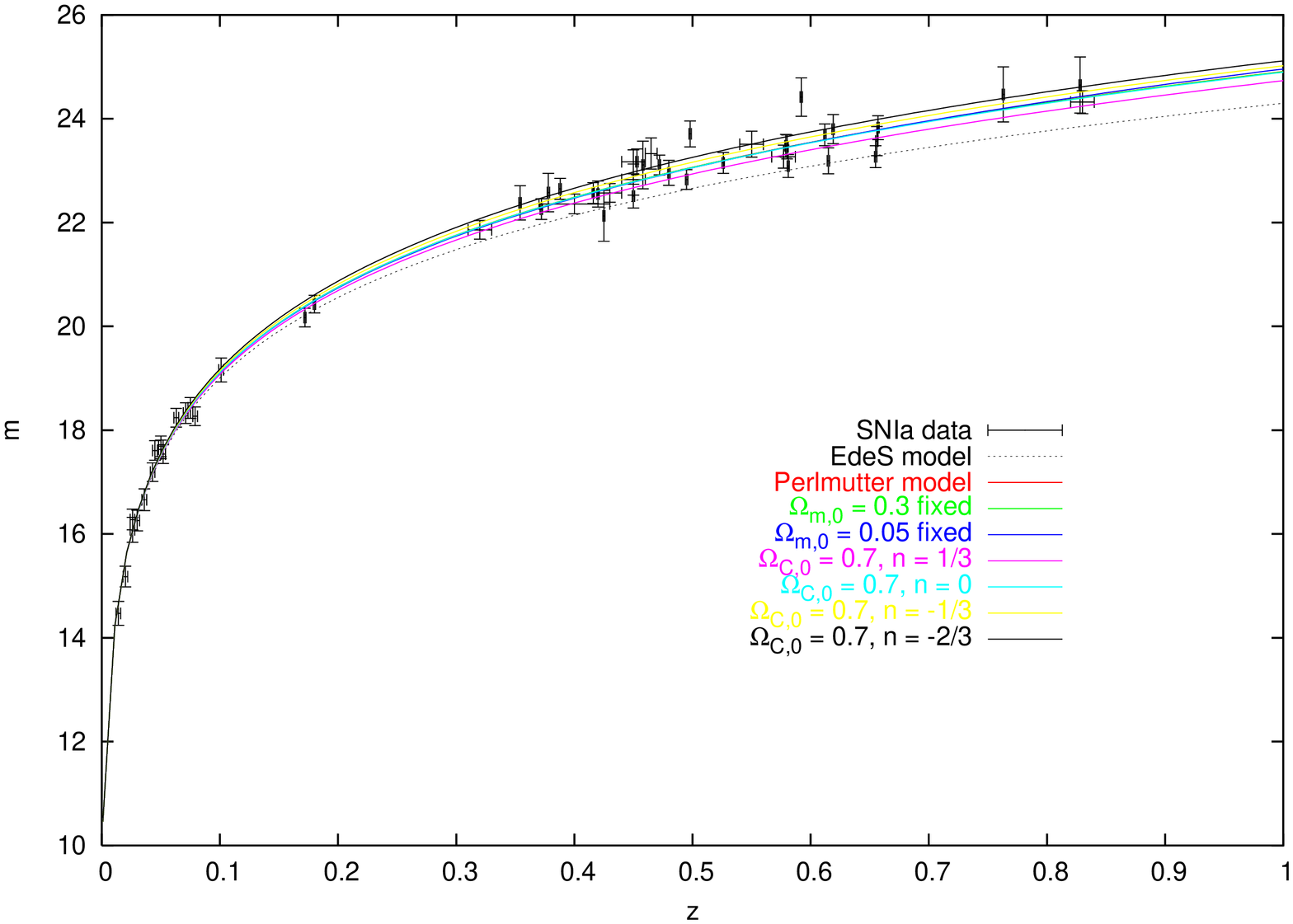} & 
\includegraphics[scale=0.3, angle=270]{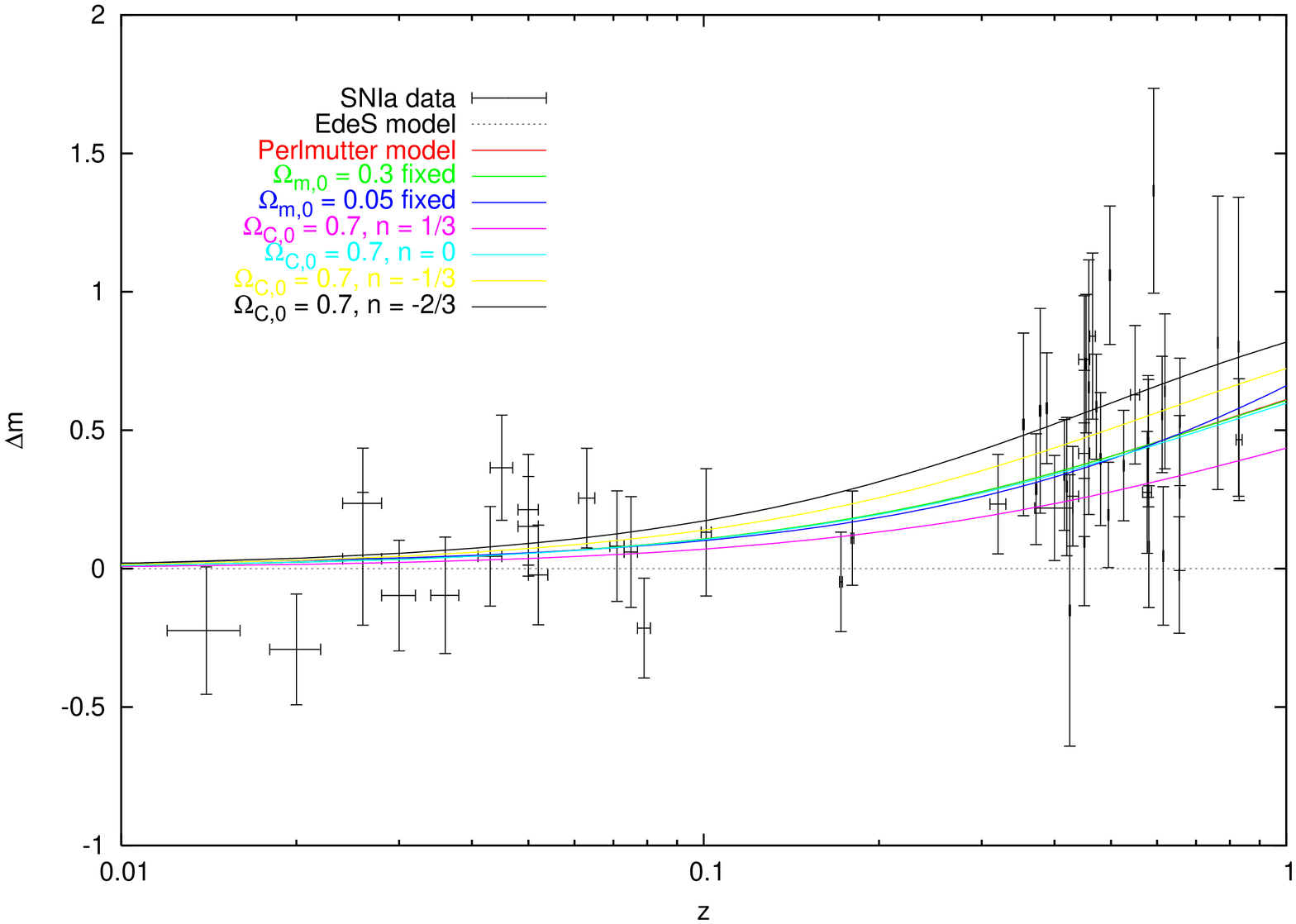} \\ [0.4cm]
\end{array}$
\end{center}
\linespread{0.5}
\caption{Hubble diagrams for Perlmutter SNIa data for sample A ({\bf a,b}) and C ({\bf c,d}). Coloured 
curves represent the best fitted magnitude-redshift relations (RGB colour standard) (The Perlmutter model 
and polynomial fits with fixed $\Omega_{m,0}$ parameters respectively) and hypothetical Cardassian scenarios
(CMYK colour standard) for different values of parameters $\Omega_{C,0}$ and $n$.
}
\label{magniacrnr}
\end{figure}

For visualization the quality of performed fitting procedure the Hubble diagrams are presented on 
Fig.~\ref{magniacrnr}.

The general conclusion from our statistical analysis is that the potential function has maximum at some 
$z=z_{0}$ (or $a=a_{0}$) and that 
\begin{equation}
\frac{\partial^{2}V}{\partial a^{2}}(z) < 0.
\label{eq:15}
\end{equation}

In the next section it will be demonstrated that the reconstructed potential function contain all 
informations which we need for reconstruction of the dynamics of any cosmological model on the phase plane.

\begin{table}[!ht]
\caption{Fitting results of the statistical analysis from distant type Ia supernovae data. The best fit
values of parameters $A_{i}$ and $\mathcal{M}$ for the polynomial fit (\ref{eq:12}) for the sample A of
Perlmutter SNIa data. In the table the values of the best fit parameters as well as the most probable
values are also given. The fit for the sample C is more robust one, because the two very likely reddened
supernovae SN1996cg and SN1996cn have been removed \cite{perlmutter99}
}
\begin{center}
\begin{ruledtabular}
\begin{tabular}{c|c|c|c|c|c|c|c|c}
  & $A_{0}$ & $A_{1}$ & $A_{2}$ & $A_{3}$ & $\mathcal{M}$ & $\chi^{2}$ & $z_{max}$ & $V_{max}$ \\
\hline
best fit & $0.61$ & $0.09$ & $0.00$ & $0.30$ & $-3.38$ & $96.03$ & $0.659$ & $-1633.21$ \\
max(P) & $0.60 ^{+0.40}_{-0.43}$ & $0.09 ^{+0.43}_{-0.40}$ & $0.00$ & $0.30$ & $-3.38 ^{+0.06}_{-0.06}$&&&\\
\hline
best fit & $-0.25$ & $1.20$ & $0.00$ & $0.05$ & $-3.37$ & $96.84$ & $3.677$ & $-1011.88$ \\
max(P) & $-0.26 ^{+0.38}_{-0.44}$ & $1.20 ^{+0.44}_{-0.38}$ & $0.00$ & $0.05$ & $-3.36 ^{+0.06}_{-0.06}$&&&\\
\end{tabular}
\end{ruledtabular}
\end{center}
\label{resultsa}
\end{table}
 
\begin{table}[!ht]
\caption{Fitting results of the statistical analysis from distant type Ia supernovae data. The best fit
values of parameters $A_{i}$ and $\mathcal{M}$ for the polynomial fit (\ref{eq:12}) for the sample C of 
Perlmutter SNIa data.}
\begin{center}
\begin{ruledtabular}
\begin{tabular}{c|c|c|c|c|c|c|c|c}
  & $A_{0}$ & $A_{1}$ & $A_{2}$ & $A_{3}$ & $\mathcal{M}$ & $\chi^{2}$ & $z_{max}$ & $V_{max}$ \\
\hline
best fit & $0.74$ & $-0.04$ & $0.00$ & $0.30$ & $-3.43$ & $53.28$ & $0.676$ & $-1567.33$ \\
max(P) & $0.73 ^{+0.40}_{-0.44}$ & $0.04 ^{+0.44}_{-0.40}$ & $0.00$ & $0.30$ & $-3.43 ^{+0.06}_{-0.06}$&&&\\
\hline
best fit & $-0.13$ & $1.08$ & $0.00$ & $0.05$ & $-3.42$ & $53.83$ & $3.522$ & $-968.74$ \\
max(P) & $-0.14 ^{+0.40}_{-0.44}$ & $1.08 ^{+0.44}_{-0.39}$ & $0.00$ & $0.05$ & $-3.41 ^{+0.06}_{-0.06}$&&&\\
\end{tabular}
\end{ruledtabular}
\end{center}
\label{resultsc}
\end{table}

\begin{table}[!ht]
\caption{Fitting results of the statistical analysis for the Perlmutter model and both (A and C) samples of 
Perlmutter SNIa data}
\begin{center}
\begin{ruledtabular}
\begin{tabular}{c|c|c|c|c|c|c|c}
  & sample & $\Omega_{\Lambda,0}$ & $\Omega_{m,0}$ & $\mathcal{M}$ & $\chi^{2}$ & $z_{max}$ & $V_{max}$ \\
\hline
best fit & C & $0.71$ & $0.29$ & $-3.43$ & $53.29$ & $0.684$ & $-1573.69$ \\
max(P)   & C & $0.70 ^{+0.05}_{-0.05}$ & $0.30 ^{+0.05}_{-0.05}$& $-3.43$ &&&\\
\hline
best fit & A & $0.69$ & $0.31$ & $-3.39$ & $96.00$ & $0.632$ & $-1628.31$ \\
max(P)   & A & $0.68 ^{+0.05}_{-0.05}$ & $0.32 ^{+0.05}_{-0.05}$& $-3.39$ &&&\\
\end{tabular}
\end{ruledtabular}
\end{center}
\label{resultspac}
\end{table}

\section{Application -- towards the reconstruction of phase portrait and model parameters}

The dynamics of the considered cosmological models is governed by the dynamical system
\begin{align}
\dot{x}&\equiv \frac{dx}{dt} =y,\nonumber \\
\dot{y}&\equiv \frac{dy}{dt} =-\frac{\partial V}{\partial x},
\label{eq:16}
\end{align}
with the first integral for (\ref{eq:16}) $\mathcal{H}=0 \Leftrightarrow \dot{x}^{2}/2+V(x)=0$.
The main aim of the dynamical system theory is the investigation of the space of all solutions (\ref{eq:16})
for all possible initial conditions, i.e. phase space $\mathbb{M}$. In the context of quintessential models 
with the equation of state $p=w(a(z))\rho$ there exists a systematic method of reducing Einstein's field 
equations to the form of the dynamical system (\ref{eq:16}) \cite{szydlowski03}. One of the features 
of such a representation of dynamics is the possibility of resolving of some cosmological problems like 
the horizon and flatness problems in terms of the potential function $V(x)$.

The phase space $\mathbb{M}$ (or state space) is a natural visualization of the dynamics of any model. 
Every point $P=(x,y) \in \mathbb{M}$ corresponds to a possible state of the system.
The r.h.s of the system (\ref{eq:16}) define a vector field 
$\mathbf{F}(P)=[y,-\frac{\partial V}{\partial x}]$ belonging to the tangent space ${\rm T_p}\mathbb{M}$.
Integral curves of this vector field define one-parameter group of diffeomorphisms $\phi(P)$ called the
phase flow. In the phase space the phase curves (orbits of the group $\phi(P)$) represent the evolution of 
the system whereas the critical points $y=0$, $\frac{\partial V}{\partial x}=0$ are singular 
solutions -- equilibria from the physical point of view. The phase curves together with critical points
constitute the phase portrait of the system.

Now we can define the equivalence relation between two phase portraits (or two vector fields) by the 
topological equivalence, namely two phase portraits are equivalent if there exists an orientation preserving 
homeomorphism transforming integral curves of both systems into each other.
Following the Hartman-Grobman theorem, near hyperbolic critical points 
($\forall$ $i$ ${\rm Re}\lambda_{i}\neq0$, where $\lambda_{i}$ is the appropriate eigenvalue of
linearization matrix $\mathcal{A}$ of the dynamical system) is equivalent to its linear part 
\begin{align}
\dot{x}&=y,\nonumber \\
\dot{y}&=-\Bigg(\frac{\partial^{2}V}{\partial x^{2}}\Bigg)_{(x=x_{0},0)}(x-x_{0}).
\label{eq:17}
\end{align}
In our case the linearization matrix takes the form
\begin{equation}
\mathcal{A} = \left[
\begin{matrix}
0,& 1\\
-\frac{\partial^{2} V}{\partial x^{2}},& 0\\
\end{matrix}
\right]_{(x_{0},0)} 
\label{eq:18}
\end{equation}

Classification of critical points is given in terms of eigenvalues of the linearization matrix since the 
eigenvalues can be determined from the characteristic equation 
$\lambda^{2}-({\rm Tr}\mathcal{A})\lambda+{\rm det}\mathcal{A}=0$. In our case ${\rm Tr}\mathcal{A}=0$ 
and eigenvalues are either real if $\frac{\partial^{2} V}{\partial x^{2}}\big|_{(x_{0},0)}<0$ or purely 
imaginary and mutually conjugated if $\frac{\partial^{2} V}{\partial x^{2}}\big|_{(x_{0},0)}>0$. 
In the former case the critical points are saddles and in the latter case they are centres.

The advantage of representing dynamics in terms of Hamiltonian (\ref{eq:1}) is the possibility to discuss
the stability of critical points which is based only on the convexity of the potential function.
In our case the only possible critical points in a finite donain of phase space are centres or saddles.

The dynamical system is said to be structurally stable if all other dynamical systems (close to it in a 
metric sense) are equivalent to it. Two-dimensional dynamical systems on compact manifolds form an open
and dense subsets in the space of all dynamical systems on the plane \cite{peixoto62}. Structurally stable
critical points on the plane are saddles, nodes and limit cycles whereas centres are structurally unstable.
There is a widespread opinion among scientists that each physically realistic models of the universe should 
possess some kind of structural stability -- because the existence of many drastically different 
mathematical models, all in agreement with observations, would be fatal for the empirical method of science
\cite{thom77}.

Basing on the reconstructed potential function one can conclude that: \\
\noindent
{\bf 1)} since the diagram of the potential function $V(a(z))$ is convex up and has a maximum which 
corresponds to a single critical point -- the quantity 
$\frac{\partial^{2} V}{\partial a^{2}}\big|_{(a_{0},0)}<0$ at the critical point (saddle point) and the 
eigenvalues of the linearization matrix at this point are real with oposite signs;\\
\noindent
{\bf 2)} the model is structurally stable, i.e., small perturbation of it do not change the structure
of the trajectories in the phase plane;\\
\noindent
{\bf 3)} since $\ddot{a}=-\frac{\partial V}{\partial a}$ one can easily conclude from the geometry of the
potential function that in the accelerating region ($\ddot{a}>0$) $V(a)$ is a decreasing function of its 
argument;\\
\noindent
{\bf 4)} the reconstructed phase portrait for the system is equivalent to the portrait of the model with 
matter and the cosmological constant.\\
\begin{figure}[!ht]
\begin{center}
\includegraphics[scale=0.45, angle=270]{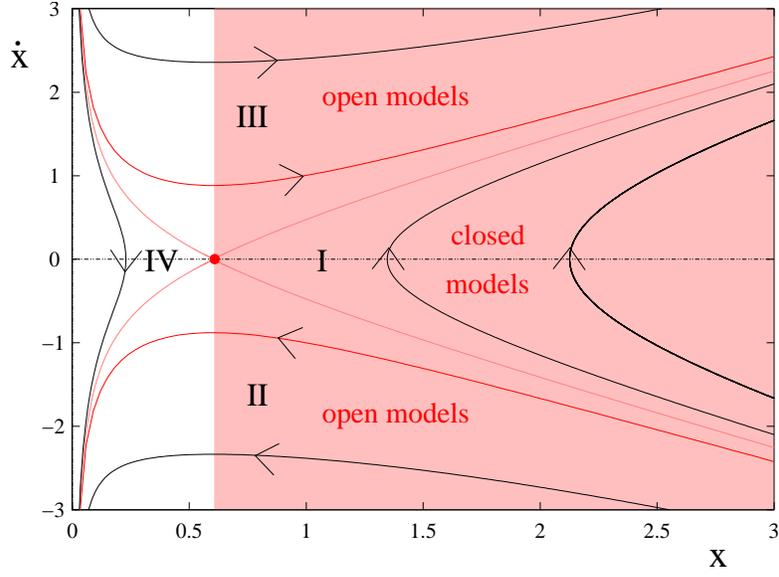} 
\caption{Phase space for the system (\ref{eq:16}) reconstructed from the potential function (\ref{eq:14})
for the best fitted parameters (TABLE \ref{resultsa}, $A_{3}=0.3$). The coloured domain of phase space 
is the domain of accelerated expansion of the universe. The red curve represents the flat model trajectory
which separates the regions with negative and positive curvature.
}
\label{figa}
\end{center}
\end{figure}
\begin{figure}[!ht]
\begin{center}
\includegraphics[scale=0.7, angle=0]{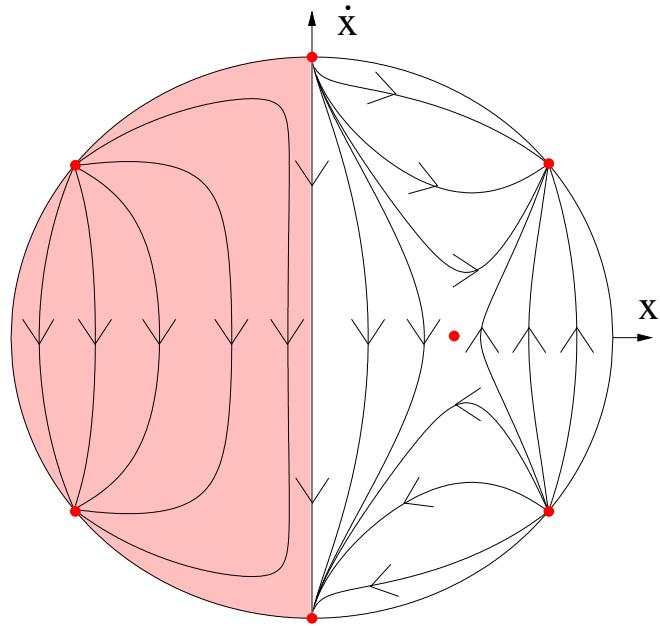}
\caption{Phase space shown in Fig.~\ref{figa} transformed on the compact Poincar{\'e} sphere. Non-physical 
domain of phase space is marked as pink.
}
\label{figap}
\end{center}
\end{figure}
By an inspection of the phase portraits we can distinguish four characteristic regions in the phase space. 
The boundaries of region I are formed by a separatrix coming out from the saddle point and going to the 
singularity and another separatrix coming out of the singularity and approaching the saddle. This region is 
covered by trajectories of closed and recolapsing models with initial and final singularity.

The trajectories moving in region IV are also confined by a separatrix and they correspond to the
closed universes contracting from the unstable de Sitter node towards the stable de~Sitter node.

The trajectories situated in region III correspond to the models expanding towards stable de Sitter
node at infinity. Similarly, the trajectories in the symmetric region II represent the universes
contracting from the unstable node towards the singularity.

The main idea of the qualitative theory of differential equations is the following: instead of finding and 
analyzing an indiwidual solution of the model one investigates a space of all possible solutions. 
Any property (for example acceleration) is believed to be realistic if it can be attributed to a large 
subsets of models within the space of all solutions or if it possesses certain stability property which is 
shared also by all slightly perturbed models.

\section{Conclusion}

The possible existence of the unknown form of matter called dark energy has usually been invoked as the 
simplest way to explain the recent observational data of SNIa. However, the effects arising from the new 
fundamental physics can also mimic the gravitational effects of dark energy through a modification of the 
Friedmann equation.

We exploited the advantages of our method to discriminate among different dark energy models. With the
independently determined density parameter of the Universe ($\Omega_{m,0}=0.3$) we found that the current
observational results require the cosmological constant $n \simeq 0$ in the Cardassian models. 
On Fig.~\ref{figpotw} we can see that both in the case of sample A (Fig.~\ref{figpotw}{\bf c}) and sample C
(Fig.~\ref{figpotw}{\bf a}) $n$ should be close to zero.
Similarly if we assume that the density parameter for barionic matter is $\Omega_{m,0}=0.05$ then 
$n \simeq 0.36$ in the case of sample C (Fig.~\ref{figpotw}{\bf b}) and $n \simeq 0.39$ for sample A 
(Fig.~\ref{figpotw}{\bf d}). Moreover, we showed (for the sample C of Perlmutter SNIa data) that a simple 
Cardassian model as a candidate for dark energy is ruled out by our analysis if 
$0.36 \lesssim n \lesssim -0.50$ for $\Omega_{m,0}=0.3$ and if $0.48 \lesssim  n \lesssim 0.13$ for 
$\Omega_{m,0}=0.05$ at the confidence level $2\sigma$.

Therefore, the standard ${\rm \Lambda CDM}$ seems to be more appropriate for a large range of model 
parameters than the dark energy models based on the first version of Cardassian models.

Our main result is that the structure of the phase space of accelerating models can be effectively 
reconstructed from SNIa data.

\begin{acknowledgments}
M. Szyd{\l}owski acknowledges the support of Jagiellonian University Rector's Scholarship.
Authors are very gratefull to dr A. Krawiec and dr W. God{\l}owski for discussion and comments.
\end{acknowledgments}

\end{document}